# *COBE* Constraints on a Local Group X-ray Halo

A. J. Banday[1] and K. M. Górski[2,3]

[1] *Max Plank Institut Für Astrophysik, 85740 Garching bei München, Germany, e-mail: banday@mpa-garching.mpg.de*
[2] *Theoretical Astrophysics Center, Juliane Maries Vej 30, 2100 Copenhagen Ø, Denmark, e-mail: gorski@tac.dk*
[3] *Warsaw University Observatory, Aleje Ujazdowskie 4, 00-478 Warszawa, Poland*

**ABSTRACT**

We investigate the effect of a putative X-ray emitting halo surrounding the Local Group of galaxies, and specifically the possible temperature anisotropies induced in the *COBE*-DMR four-year sky maps by an associated Sunyaev-Zel'dovich effect. By fitting the isothermal spherical halo model proposed by Suto et al. (1996) to the coadded four-year *COBE*-DMR 53 and 90 GHz sky maps in Galactic coordinates, we find no significant evidence of a contribution. We therefore reject the claim that such a halo can affect the estimation of the primordial spectral index and amplitude of density perturbations as inferred from the DMR data. We find that correlation with the DMR data imposes constraints on the plausible contribution of such an X-ray emitting halo to a distortion in the CMB spectrum (as specified by the Compton-$y$ parameter), up to a value for $R$ – the ratio of the core radius of the isothermal halo gas distribution to the distance to the Local Group centroid – of 0.68. For larger values of $R$, the recent cosmological upper limit derived by *COBE*-FIRAS provides stronger constraints on the model parameters. Over the entire parameter space for $R$, we find an upper limit to the inferred sky-RMS anisotropy signal of $\sim 14$ $\mu$K (95% confidence), a negligible amount relative to the $\sim 35$ $\mu$K signal observed in the *COBE*-DMR data.

## 1 INTRODUCTION

The initial detection of large angular scale anisotropy in the cosmic microwave background radiation (CMB) by the *COBE*-DMR instrument (Smoot et al. 1992), and its subsequent confirmation with four years of data (Bennett et al. 1996), affords strong support to theories which posit that the structure observed in the universe today evolved via gravitational instability from seed fluctuations generated during an inflationary phase in the early universe. This issue is conveniently pursued in terms of the power spectrum of the signal in the four-year sky maps, which provide a comprehensive data set against which the predictions of theories for cosmological structure formation can be tested. Górski et al. (1996a) have investigated the data within the framework of primordial density perturbations with power law spatial perturbations, $P(k) \propto k^n$. The normalisation is conveniently parameterised by the exact value of the quadrupole amplitude, $Q_{rms-PS}$, as introduced by Smoot et al. (1992). Within this class of models, the four-year data can be adequately described by a Harrison-Zel'dovich spectrum ($n = 1$) with $Q_{rms-PS} \sim 18$ $\mu$K. Considerable investigation of more specific cosmological models has been undertaken with the two-year data (e.g. see Stompor, Górski & Banday 1995) and the four-year sky maps are currently undergoing a similar degree of scrutiny (see Stompor, 1996; Górski et al., 1996b; Bunn & White, 1996 for example).

An important issue in this context which has been addressed is the extent to which local astrophysical foregrounds can systematically contaminate the detected signal. Comprehensive investigations (Banday et al. 1996; Górski et al. 1996a; Kogut et al. 1996) demonstrate that known foregrounds result in *minimal* corrections to the dominant CMB signal. In particular, Banday et al. (1996) show that the combined effect of Galactic and extragalactic signal subtraction on the most likely ($Q_{rms-PS}$, $n$) is to shift these values insignificantly by $\sim 0.1$ $\mu$K and -0.06 relative to their uncorrected values. Of course, such limits only apply in the context of the foregrounds against which the DMR observations were compared, and it is useful to place limits on other potential, but inevitably small, non-cosmological signals in the data.

Suto et al. (1996) have recently proposed a model in which the Local Group (LG) of galaxies has its own X-ray halo. Such a halo could generate anisotropies in the CMB via the Sunyaev-Zel'dovich (SZ) effect. The original motivation for suggesting such a contribution may be found in the possibility of explaining the 'low' quadrupole observed in the DMR two-year data. In fact, their simple analytic calculation implies that the halo-induced quadrupole can be comparable to the observed sky quadrupole without violating limits on the Compton $y$-parameter measured by the *COBE*-FIRAS instrument. However, Kogut et al. (1996) have demonstrated that the cosmic quadrupole is counter-aligned to the Galactic quadrupole, and determination of its amplitude is dominated by uncertainties in the Galactic modelling. The quadrupole as observed in DMR sky maps uncorrected for Galactic emission is not representative of the CMB quadrupole, which is constrained to lie in the range [4, 28] $\mu$K. Despite the apparent redundancy of the initial motivation, investigation of the model remains interesting.

In this paper, we reconsider the halo model and place limits on its possible contributions to the four-year DMR sky maps using the correlation technique described in detail in Górski et al. (1996a) and implemented in Banday et al. (1996).

## 2 ANALYSIS TECHNIQUE

### 2.1 Data Selection

We utilise the DMR four-year 53 and 90 GHz full sky maps in Galactic coordinates (Bennett et al. 1996). The four sky maps (two at each frequency) are coadded using inverse-noise-variance weights. We do not use the 31 GHz data since under such a weighting scheme the contribution of these least sensitive DMR maps is minimal.

The dominant contribution to the sky signal at microwave frequencies is from structure associated with the Galactic plane. Since it can not be modelled to sufficient accuracy to enable its subtraction from the data, we excise those pixels where the CMB anisotropy is necessarily dominated by such emission. We employ a cut which eliminates all pixels for which the Galactic latitude is less than 20° (as in previous DMR analyses), and further bright regions – as traced by the $DIRBE$-140 $\mu$m sky map – in Scorpius-Ophiuchus and Taurus-Orion (Banday et al. 1997). The new cut leaves 3881 surviving pixels.

### 2.2 Temperature Anisotropy due to the Local Group Halo

Suto et al. (1996) provide a formula for the spatial variation of the SZ temperature decrement associated with a spherical isothermal plasma (their eqs 2 and 4) which we rewrite as,

$$\Delta T(\mu) \propto \frac{R^2}{\sqrt{1+R^2-\mu^2}} [\frac{\pi}{2} + sin^{-1}(\frac{\mu}{\sqrt{1+R^2}})]$$

where we have written $R = \frac{r_c}{x_0}$. The observed temperature along a given line-of-sight specified in Galactic coordinates thus depends on the cosine of the angle ($\mu$) between this direction and the LG halo centroid. Given the angular coordinates of the centroid and a value for the ratio ($R$) of the gas core radius ($r_c$) to the centroid distance from the Galaxy ($x_0$), the predicted spatial morphology is unique (although its overall normalisation is not) and it is simple to generate a map of the expected anisotropy. Since the core radius of the LG gas distribution is unknown, we generate maps of the temperature distribution for a number of values of $R$. Each map is convolved with the $COBE$-DMR beam.

### 2.3 Correlation Methodology

As an initial introduction to the technique, consider the simple case in which we cross-correlate the DMR maps with a given sky map $X$ over the high latitude sky as defined previously. Initially, the DMR sky map and map $X$ are decomposed into Fourier coefficients in a basis of orthonormal functions which specifically includes both pixelisation effects and the Galactic cut (Górski 1994). This technique has the advantage of allowing exact exclusion of the monopole and dipole moments (which are uninteresting for the investigation of primordial anisotropy) from the analysis, and makes full use of all of the available spatial (phase) information. The measured DMR map Fourier coefficients, $\mathbf{c}_{DMR}$, can then be written in vector form as $\mathbf{c}_{DMR} = \mathbf{c}_{CMB} + \mathbf{c}_{N} + \alpha_X \mathbf{c}_X$ where $\mathbf{c}_{CMB}$, $\mathbf{c}_{N}$ and $\mathbf{c}_X$ are the coefficients for the CMB anisotropy, the noise and the external map. $\alpha_X$ is now a coupling constant (with units $\mu$K $X^{-1}$) with exact solution (Banday et al. 1996)

$$\alpha_X = \mathbf{c}_X \tilde{M}^{-1} \mathbf{c}_{DMR} / \mathbf{c}_X \tilde{M}^{-1} \mathbf{c}_X$$

where $\tilde{M}$ is the covariance matrix describing the correlation between different Fourier modes on the cut-sky and is dependent on assumed CMB model parameters and the DMR instrument noise. A full-description of its derivation is given in Górski (1994). The contribution of $X$ to the DMR sky is the map $\alpha_X T_X$, where $T_X$ is the 'temperature' of the foreground emission in map $X$ in the appropriate natural units. Fig.1 shows the power spectra computed for the LG halo model on the cut-sky for several values of $R$. These arbitrarily normalised spectra can be compared to the observed power spectrum of the coadded DMR 53 and 90 GHz sky map.

High-latitude Galactic emission can be at least partially accounted for by using the $DIRBE$-140 $\mu$m sky map (Reach et al. 1995) as a template for free-free and dust emission, and a full-sky radio survey at 408 MHz (Haslam et al. 1981) to trace the synchrotron contribution. We extend the above formalism (as described fully in Górski et al. 1996a) to allow us to determine the Galactic contribution by fitting the above two templates simultaneously with the LG halo sky map to the DMR data.

The covariance matrix requires a prescription for the CMB. Although Suto et al. have suggested that the effect of the LG halo will have ramifications for the determination of spectral parameters from the DMR data, we elect to describe the cosmic signal in terms of a Harrison-Zel'dovich power law spectrum with RMS quadrupole normalisation $Q_{rms-PS} = 18$ $\mu$K as suggested by the recent DMR analysis (Górski et al. 1996a). We have investigated the robustness of our calculations to this assumption by utilising a variety of alternative cosmological anisotropy power spectra, and this is discussed later in the paper.

### 2.4 Results

The likelihood function for the DMR sky map as corrected by two Galactic templates and the LG halo model parameterised as a function of $R$ is shown in Fig. 2. The likelihood ratio between the DMR data corrected solely by the Galactic templates and with a best-fit halo contribution removed indicates that we are unable to exclude any value for $R$ in the interval 0 - 10. The only significant correction to the DMR Fourier coefficients comes from the observed correlation with the $DIRBE$-140 $\mu$m sky map. The SZ effect due to a LG halo does not affect any determinations of cosmological parameters as inferred by the DMR sky maps.

We have computed the coupling constant and its associated error between the halo template and the DMR sky map as a function of $R$, Since there is no statistically significant detection of correlation, we establish limits on the LG model contribution by scaling the temperature distribution by twice the coupling error. The mean scaled temperature, $\alpha_X \overline{T_X}$, of the halo template can be related to the Compton-$y$ parameter through the relation $y = \alpha_X \overline{T_X}/(2T_{cmb})$. Recent *COBE*-FIRAS measurements (Fixsen et al. 1996) set an upper limit on the Compton-$y$ distortion of $15 \times 10^{-6}$. We note that for values of $R$ over the range [0.,0.68], the DMR imposed limits provide the strongest constraints on the model. Beyond this value of $R$, FIRAS provides tighter limits. This should not be too surprising given the changes in the halo spatial distribution with increasing $R$. At small values of the parameter, the spatial distribution is dominated by a cold spot centered in angular coordinates at M31, and there is significant structure ove a range of angular scales, enabling DMR to provide significant constraints on the model contribution to the sky anisotropy. As $R$ increases, however, the distribution of power is progressively dominated by the largest angular scales, particularly the monopole and dipole which DMR is insensitive to (as descibed above). For the values of $r_c$ and $x_0$ (ie. $R \sim 0.43$) from Suto et al. (1996), DMR constrains the LG halo to generate at most of order 50% of the observed limit on the Compton-$y$ distortion from FIRAS. Over the entire range of $R$, we establish upper limits to the quadrupole amplitude and sky-RMS as measured by DMR (i.e. on the cut sky and after best-fit monopole and dipole subtraction) of $\sim 13\,\mu$K and $\sim 14\,\mu$K respectively (at 95% c.l.). Given that these are upper limits, this is insignificant relative to the observed RMS anisotropy in the DMR sky maps of $\sim 35 \pm 2$ $\mu$K (Banday et al. 1997). Although the upper limit on the quadrupole contribution would appear to be rather large, the cosmological amplitude is already rather poorly determined as a consequence of Galactic foregrounds. Most importantly, the limit on the halo quadrupole is not sufficiently large to perturb any conclusions inferred from *COBE*-DMR about cosmological anisotropy.

The sensitivity of our results to assumptions about the power spectrum representing the dominant cosmological part of the anisotropy measured by DMR has been investigated. For the class of models described by power law spectra, we repeated our computations with $n = 0.6$ and 1.7 and the appropriate normalisations from Górski et al. (1996a). There is *no* statistically significant change in the correlation of the halo model with DMR. For anisotropy models where the quadrupole is significantly suppressed (e.g. open models) or excessive (e.g. low-density spatially flat models) relative to the $n = 1$ power law model, we find variations in our limits of order $\pm 20\%$. However, those models where the deviations are largest are somewhat disfavoured by a combination of the DMR data with additional cosmological constraints (see Górski et al., 1996b, for a summary of such constraints). We conclude that our results are fairly robust.

## 3  DISCUSSION

Very recently, Pildis & McGaugh (1996) have concluded, by comparison to recent observations of poor groups of galaxies by the *ROSAT* PSPC, that any Local Group halo will be too tenuous to produce the effects that Suto et al. (1996) discuss. In this work, we have demonstrated that the observed anisotropy in the *COBE*-DMR data does not support the existence of SZ anisotropy due to a Local Group X-ray halo. Furthermore, although only weak constraints can be imposed on the parameters of the model from our analysis, we have shown that, irrespective of any assumptions or extrapolations about the nature of the putative X-ray halo, limits on an associated SZ effect and particularly its anisotropy imply that *no* significant modifications to fits of cosmological power spectra are allowed.

The dominant feature in such a halo model, a cold spot towards M31 is partially excluded from the fit by virtue of its proximity to the Galactic cut applied to the DMR data, as is any information from the dipole anisotropy (which is assumed to have a Doppler origin in the CMB data), and this weakens the model limits which our analysis can furnish. Improved limits may be possible by direct comparison of the model with X-ray data, particularly that measured by *ROSAT*, where the dipole information may be useful. Departures from the spherical isothermal model could modify our results, but such deviations are fairly unlikely given the dynamical properties of the LG.

Previous work (Banday et al. 1996) showed little evidence of an SZ contribution to the DMR sky as traced by the *HEAO*-A2 X-ray data, which measures photon energies in the 2 - 60 keV band. Correlation of the DMR sky maps with the well-defined spectral bands of the *ROSAT* X-ray data may be more appropriate, and analysis by Kneissl et al. (1996) shows little correlation between the 53 and 90 GHz DMR sky maps and the *R6 ROSAT* band. The *COBE*-DMR data are a clean data set with which to investigate cosmological models for CMB anisotropy on large-angular scales.

## ACKNOWLEDGMENTS

We acknowledge the efforts of those contributing to the *COBE*-DMR. We thank R. Kneissl and G. Boerner for useful comments. KMG was supported by the Danish National Research Foundation through its establishment of the Theoretical Astrophysics Center.

**FIGURE CAPTIONS**

**Figure 1.** The rms power spectra computed on the DMR cut-sky (Banday et al., 1997), normalised relative to a Harrison-Zel'dovich spectrum. The solid circles show the values for the coadded DMR 53 and 90 GHz sky map in $\mu$K. The open circles show spectra for the Local Group isothermal halo with $R$, the ratio of the core radius, $r_c$, of the isothermal X-ray emitting gas to the distance, $x_0$, of the Local Group gas centroid taking the values 0.01,0.1,0.4 and 1.0 descending down the figure. The halo spectra are arbitrarily normalised to unity at $l = 2$.

**Figure 2.** The likelihood curve (thick solid line) for the DMR coadded 53 and 90 GHz sky map, corrected for best fit contributions from two Galactic foreground templates (see the text) and the LG halo model template, as a function of $R$, the ratio of the core radius, $r_c$, of the isothermal X-ray emitting gas to the distance, $x_0$, of the Local Group gas centroid. We show the most interesting range of $R \in [0, 1]$. The solid point at $R = 0$ corresponds to the relative likelihood amplitude with *no* halo contribution (i.e. the DMR data is corrected by the Galactic model only). No constraints can be imposed on this parameter from the likelihood alone. The LG halo model shows no statistically significant correlation with the DMR data. The vertical solid line at $R = 0.68$ delineates the region to the left where DMR imposes the most significant constraints on the halo model to the righthand region where FIRAS provides stronger constraints. This dividing region lies close to the point on the likelihood curve where a LG halo contribution provides no improvement to the DMR data over a purely Galactic correction (cf. the dashed line).

**Figure 3.** The predicted sky signal (95% c.l. upper limits as a function of $R$) due to an SZ effect induced by a putative X-ray halo of the Local Group of galaxies. The Compton-$y$ parameter is shown as the solid line, the quadrupole in $\mu$K as the long dashed line, and the sky-RMS in $\mu$K as the short dashed line. The FIRAS limit on $y$ (Fixsen et al. 1996) is shown as the dotted line. Up to a value for $R$ of 0.68, note that limits imposed on the x-ray halo from DMR provide the strongest constraint, beyond this value the FIRAS upper limit is the srongest constraint. The lines shown in the heaviest type represent the upper limits on various parameters by combining DMR and FIRAS imposed limits. Lighter type shows DMR alone imposed limits.

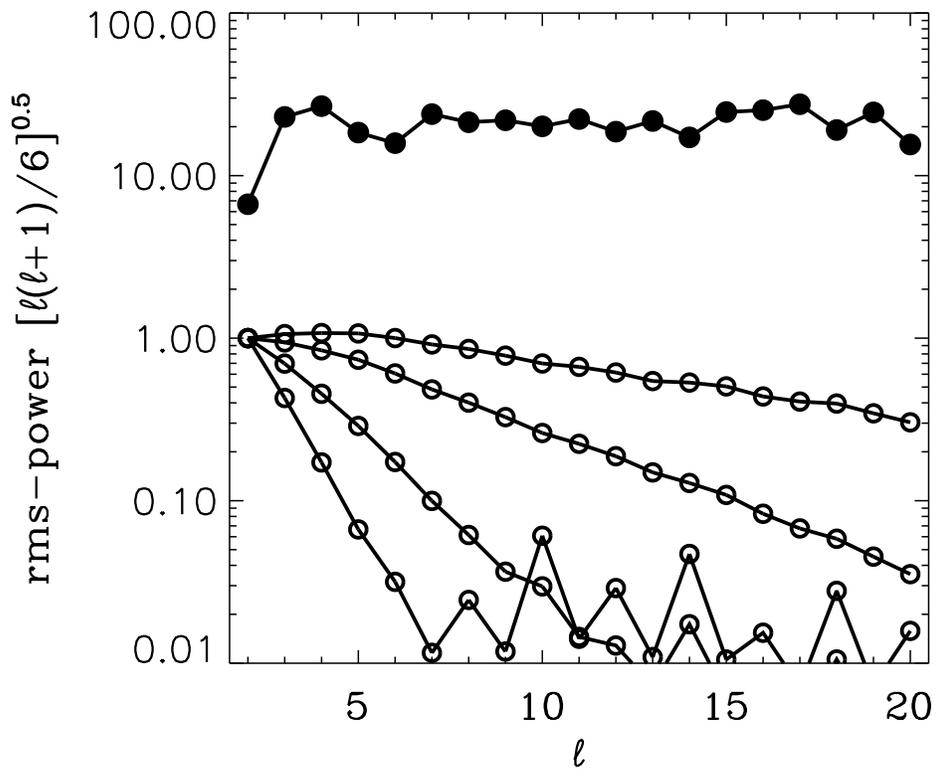

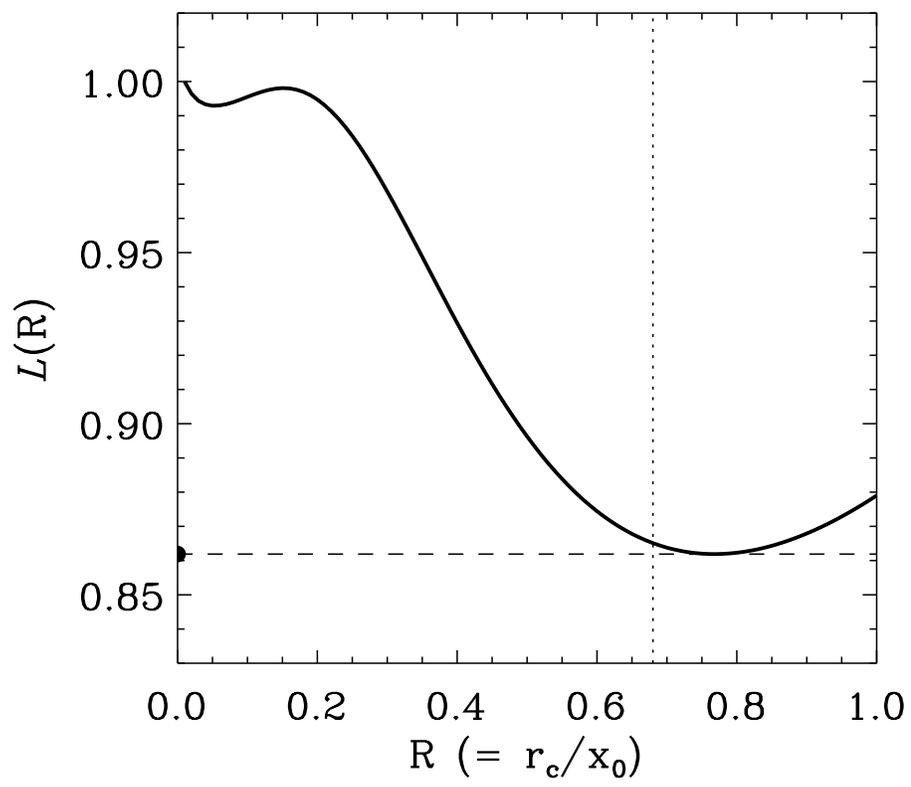

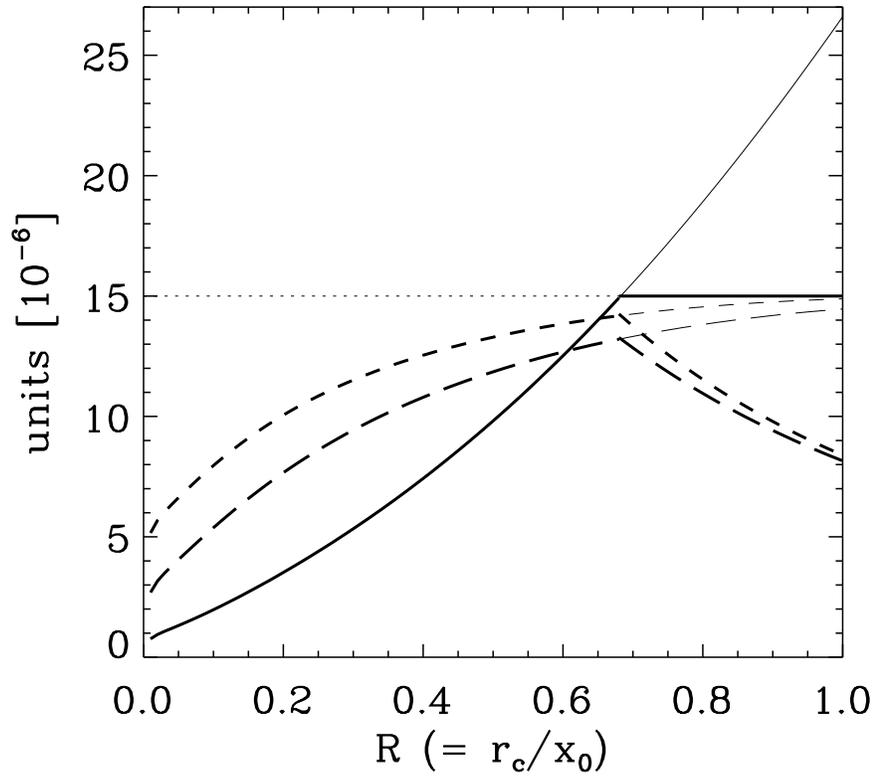